# Traveling waves induced by sweeping flows on solidification interfaces

Alain POCHEAU[1, a *], Tania JIANG[1,b] and Marc GEORGELIN[1,c]

[1]Aix-Marseille Univ, CNRS, Centrale Marseille, IRPHE, Marseille, France

[a]alain.pocheau@univ-amu.fr, [b]tetyana.jiang@univ-amu.fr, [c]marc.georgelin@univ-amu.fr

**Abstract.** Solidification of alloys in a thermal gradient usually involves the generation of flows by thermal or thermosolutal convection. We experimentally study their effects on the dynamics of a solidification interface by inducing a controlled sweeping flow in a directional solidification device. Flow is induced in the sample from an external thermosiphon. Downstream inclination of microstructures and downstream sidebranch development are observed. However, the major outcome is the evidence of large scale travelling waves on the solidification interface. They are induced by the coupling between solidification and flow and yield repetitive striations of the solid phase. Two waves are observed and characterized.

**Introduction**

Solidification of melts in large volumes usually involves flows in the liquid phase. They are generated either by thermoconvective or thermosolutal instabilities [1] or forced by seed rotation in the Czochralski process [2] or by magnetic stirring [3]. These flows then participate to the dynamics in two ways : (i) by inducing a large scale homogenization of the melt ; (ii) by coupling their solute advection to the interface dynamics. Whereas the former effect is largely taken into account in models and simulations, the latter effect has been poorly addressed so far, especially on microstructures. Here, we design a directional solidification experiment in thin samples to study it in well controlled conditions and with in-situ real-time visualization [4,5].

After describing the experimental set-up, we shall report the effects of flows on microstructures and on the interface dynamics. In particular, two interfacial waves induced by the flow will be evidenced for the first time with implications involving a repetitive striation of the solid phase.

**Experimental Set-Up**

A 15 cm x 5 cm thin sample is sandwiched between heaters and coolers and pushed by a translation stage so as to solidify at a controlled velocity V (Fig. 1a). To allow solidification at convenient temperatures, a dilute alloy of a plastic crystal (succinonitrile) is used with a melting temperature of 56°C, a solutal diffusivity D of 1350 $\mu m^2.s^{-1}$ and a partition coefficient K of 0.29. Its physical properties enables it to mimic the solidification of metals, as confirmed by numerous studies over three decades. It is studied in a thin sample of 150 μm depth, so as to involve a single microstructure layer for further visualization. Experimental and numerical studies [6] showed that, for depths above 10 μm, this layer displays the same dynamics as in unconfined samples. Samples were prepared in a single crystal state with axes aligned on the heat flow and on the sample depth.

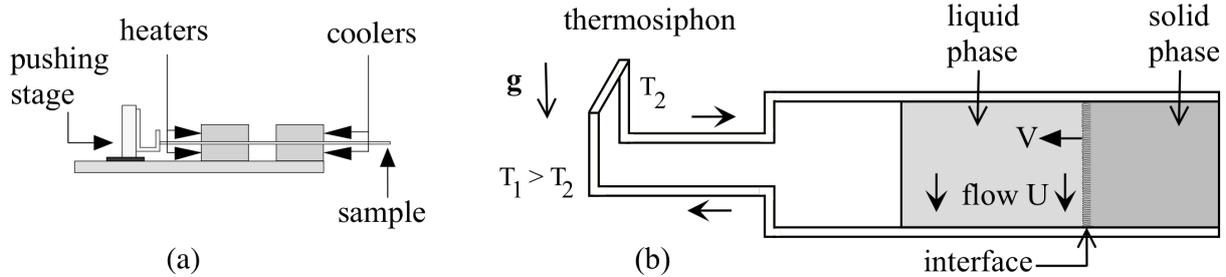

Figure 1 : Sketch of the experimental set-up showing a sample pushed so as to solidify in a thermal gradient induced by heaters and coolers. (a) Directional set-up. (b) Thermosiphon circuit added to generate a sweeping flow U parallel to the solidification interface. The growth velocity of the interface is labeled V.

All devices of this directional, Bridgman-growth, set-up are electronically regulated to better than 0.1 µm.s$^{-1}$ and 0.1°C and with extreme temperatures of 100°C and 10°C. A few millimeter gap between heaters and coolers allows real-time visualization of the solidification interface onto a camera with thermal gradients of 70 or 140 K.cm$^{-1}$. Solidification velocities can go up to 50 µm.s$^{-1}$.

This set-up, which is described in detail in [7-11], a priori involves no flows since these are prohibited in thin samples due to large viscous dissipation. This however provided the opportunity to complete it with externally generated controlled flows. For this, we let the melt flow in a closed circuit made by the liquid phase, the sample sides and a thermosiphon (Fig. 1b). Thermosiphon was chosen since, being a thermal device, it is known to stand as a reliable and constant source of micrometric flows [12]. Monitoring the temperatures of the 5 cm branches of the thermosiphon between 65°C and 97°C thus led the generation of controlled flows from one side of the sample to the other with an amplitude up to 1500 µm.s$^{-1}$.

**Flow Effects on Interface and Microstructures**

We report the effects of flows on both the solidification interface and its microstructures, for increasing velocity V.

**Inclinations and sidebranching.**

At low enough velocity V and flow U, both an inclination of cells and dendrites and an asymmetric sidebranching are observed. Both the direction of inclination and the side of sidebranch development are downstream (Figs. 2a, 2b). The inclination angle appears to raise linearly with the ratio U/V (Fig. 2c). It thus appears to be linked to the kinematic angle β induced by translation at velocity U during growth at velocity V, but with a smaller prefactor : α≈0.18 U/V.

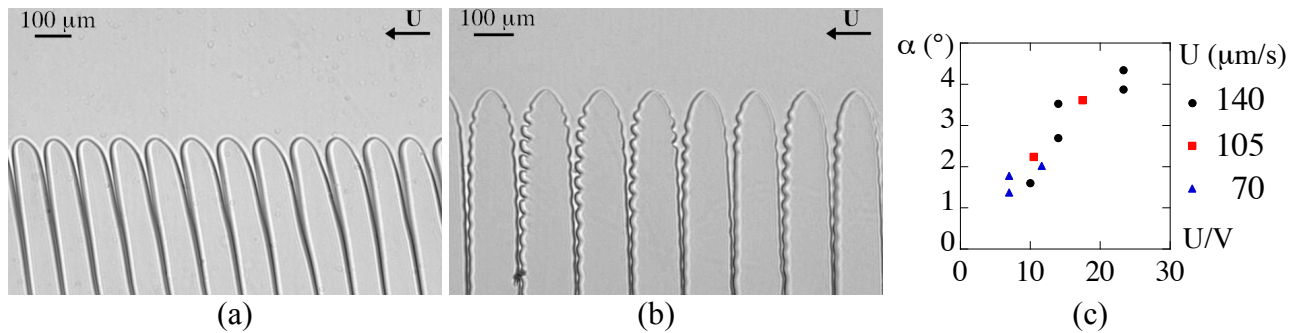

Figure 2 : Cell inclination and dendrite asymmetry induced by the flow. U=140 µm s$^{-1}$. (a) Cells with downstream inclination of α=5.8°, G=140 K cm$^{-1}$, V=6 µm s$^{-1}$. (b) Dendrites with downstream inclination of α=1.2°, G=70 K cm$^{-1}$, V=18 µm s$^{-1}$. (c) Evolution of the downstream inclination angle α with the ratio U/V.

**Waves.**

Beyond some velocity V or some flow amplitude U, a spectacular and original phenomenon occurs in the form of wavy deformations of the solidification interface. Two different kinds of waves are observed that differ by their velocity C and their form : a slow wave with smooth, quasi-sinusoidal modulation (Figs. 3a, 4a) ; a rapid wave with a sharp asymmetric profile (Figs. 3b, 4b). At large flow, the local growth conditions at the head of the rapid wave even yield a dendrite to grow parallel to the flow and thus normally to the heat flow direction (Figs. 3c, 4c).

Both waves involve a wavelength extended over many cells. As the wave passes by, cells or dendrites move back and forth on the heat flow direction and distort. Their dynamics then imprints the solid phase with permanent striations which represent a modulation of the microsegregation of the material. We report below the main features of these waves and of their striations.

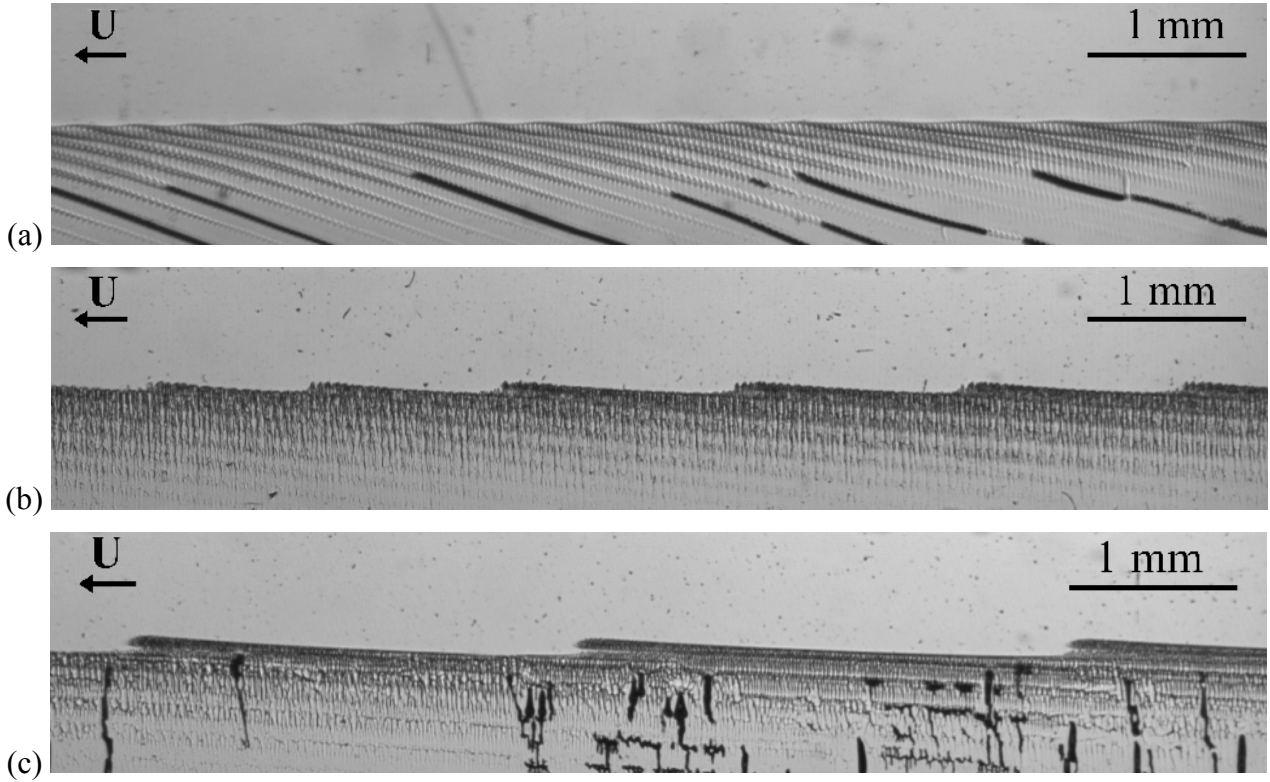

Figure 3 : Interfacial traveling waves induced by the flow. Flow intensity U=106 µm s$^{-1}$, sample depth e= 150 µm. (a) Slow wave, G=70 K cm$^{-1}$, U=106 µm s$^{-1}$, V=6 µm s$^{-1}$. (b) Rapid wave, G=140 K cm$^{-1}$, U=106 µm s$^{-1}$, V=16 µm s$^{-1}$. (c) Dendrite growing in the flow direction, normally to the heat flow : G=140 K cm$^{-1}$, U=306 µm s$^{-1}$, V=8 µm s$^{-1}$. Black channels result from a gaseous phase in the grooves which does not couple to the interface dynamics.

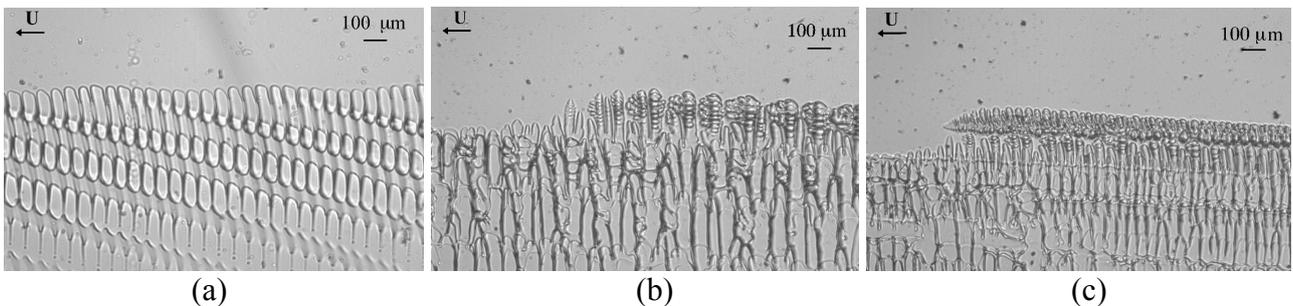

Figure 4 : Zoom of figure 3 showing the detail of the modulations of the interface and of the solid phase. (a) Slow wave. (b) Rapid wave. (c) Dendrite growing in the downstream direction.

**Analysis.**

**Phase diagram.**

Figure 5 reports the phase diagram of waves. Both waves appear unbounded on flow amplitude U but bounded on velocity V up to a value that depends on U. Slow waves are confined to small velocities V and rapid waves to intermediate velocities V. A domain of coexistence of the waves on the same interface is noticeable. It evidences the intrinsic difference between slow and rapid waves.

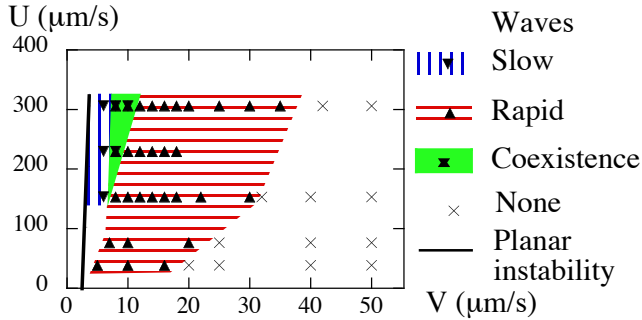

Figure 5: Phase diagram of waves in the plane (U,V).

**Wave velocities and wavelengths.**

Wave velocities C differ both in magnitude and in the way they vary with V and U. For both waves, they display no noticeable variations with V. However, rapid waves show a velocity that linearly increases with U (C≈0.4 U) whereas slow waves exhibit a nearly constant and small velocity of about 40 μm.s$^{-1}$ (Fig. 6a). Despite these differences, the two waves share the same evolution of wavelength which raises linearly with C/V (Fig. 6b). The difference of magnitude of C for the two waves justify their name.

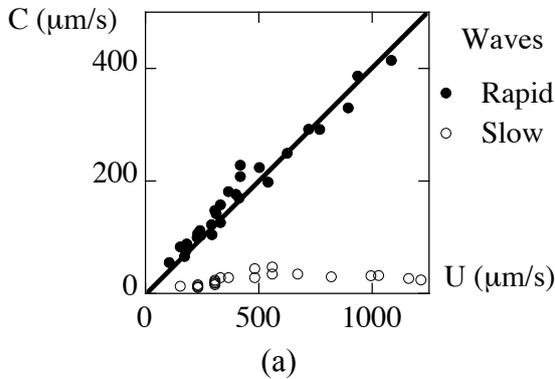 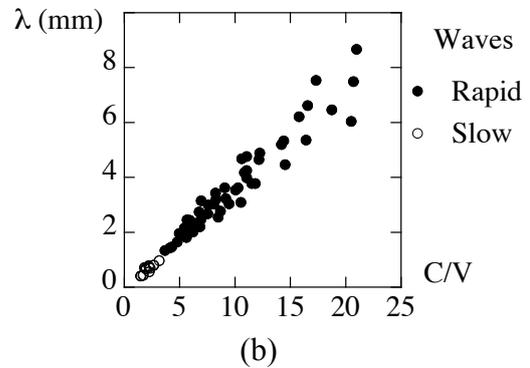

Figure 6 : For both rapid and slow waves : (a) Wave velocities C with respect to flow intensity U. (b) Wavelengths λ with respect to the ratio C/V.

**Periods, striation distances and inclinations.**

The periods T of the waves depend on both the growth velocity V and the flow amplitude U. However, they surprisingly share the same values T(V,U) despite their intrinsic differences. Interestingly, on the range of V and U studied, the length δ=VT proves to be nearly constant to about 350 μm (Fig. 7). It enables to recover the wavelength evolution since λ=CT= δ C/V.

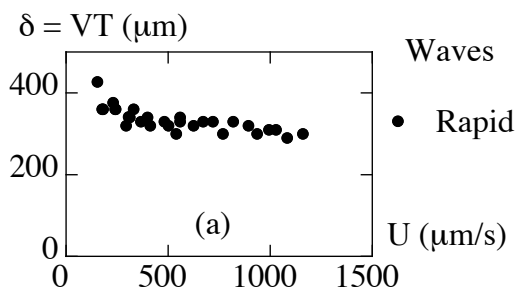 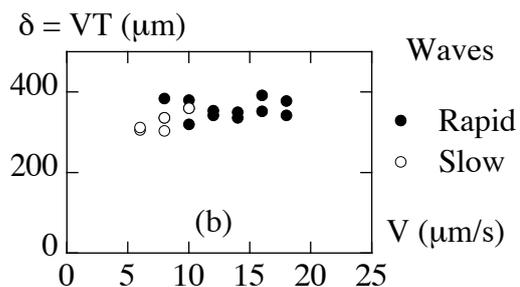

Figure 7 : Distance δ between striation lines along the heat flow direction with respect to flow intensity U (a) and the growth velocity V (b).

The length δ corresponds to the distance, in the solid phase and on the pushing (or heat flow) direction, between consecutive wave iso-phase lines (Fig. 8). The actual thickness d of the resulting bands follows from the inclination angle θ : d= δ cos(θ) where, by simple kinematics, tan(θ)=V/C.

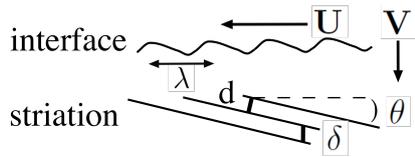

Figure 8 : Sketch of a wavy interface and the resulting modulations in the solid phase. They display an inclination angle θ, a distance δ on the heat flow direction and a band thickness d.

**Solute depletion and propagation mechanism.**

As the interface satisfies local thermodynamic equilibrium, its temperature $T_I$ follows its concentration $c_I$ according to the Gibbs-Thomson relation $T_I=T_M + m\, c_I$ where m (negative here) denotes the liquidus slope and $T_M$ the melting temperature of pure melt (the surface tension being neglected). As the thermal gradient is imposed, $T_I$ and thus $c_I$ can thus be determined along the wave profile. Figure 9 then displays on a rapid wave the solute depletion that grows along its leading edge. In addition, the wave profile $z(x,t)=z(x+Ct)$ can be related to the interface growth velocity $V_I$ in the liquid frame since $V_I = V+\partial z/\partial t = V+C\partial z/\partial x$. Its negative slope then shows that dendrites stop growing and thus reject no solute flux. Altogether, the liquid phase is thus poor in solute on this leading edge. Being advected by the flow, it then yields a large solute depletion in front of the wave head, which makes dendrites quickly grows there. This yields the wave to advance in the flow direction.

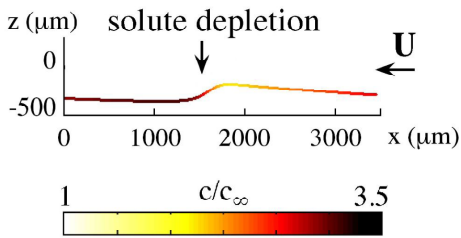

Figure 9 : Solute evolution along the interface determined from the Gibbs-Thomson relation with neglected surface tension, according to the color scale.

**Discussion.**

The externally generated flow enters from one side of the sample and sweeps the interface. A relevant question refers to its penetration length in the microstructure. If the flow hardly penetrates behind the microstructure tips, its main effect is to advect the solutal field. Advection in the downstream direction of the poor-solute melt ahead of the microstructure tips then promotes downstream inclination and sidebranching. In contrast, if the flow penetrates sufficiently far behind microstructure tips, it induces a compression of iso-concentration lines of solute on the upstream sides of microstructure. This then promotes upstream inclination and sidebranching. In the literature, the two kinds of directions have been found : upstream direction in thin sample [13-15] or in volume [16] ; downstream direction [17] or both directions [18] in thin samples. Here, as only downstream effects have been evidenced, the flow presumably induces only solute advection ahead of microstructure tips.

As the present experiment involves thin samples, one may question the relevance of its results to less confined domains. Between both, the major difference refers to the flow structure, a Poiseuille flow here versus a constant flow in large volumes. This should result in a weaker flow on the microstructure sides in thin sample than in large domains. Although this would yield the flow to penetrate deeper inside the microstructure tips in large volumes, the major effect should remain the solute advection ahead of them. To address this issue, we have increased the sample depths up to a factor 3 and found no noticeable modification of inclinations and waves, qualitatively and quantitatively regarding the wave period and celerity. In particular, the constant length δ=VT did not vary with the sample depth. As we also evidenced that it negligibly varies with the thermal gradient, it thus appears as a constant of the material. These findings give confidence that the effects observed here should remain valid in large volume, although they are far less easy to evidence directly. Their clearest signature would thus presumably be some striation which could be evidenced by post-mortem analysis. Then, the distance d between bands should be linked to V, U and the constant length δ, as described above.

Regarding materials, the major implication of the present findings is the repetitive striations induced by interfacial waves. They should provide large scale structurations of the solid phase that may be controlled or provoked by the flow magnitude.

**Conclusion.**

Adding a controlled sweeping flow to directional solidification in thin samples, definite implications of flows on microstructures and interface dynamics have been evidenced. Among them, the most salient phenomenon appears to be the generation of interfacial waves which largely modify the form and the solute rejection of microstructures, yielding repetitive striations in the solid phase. Although evidenced in thin samples, this phenomena should persist in large volume and yield repetitive banding in materials.